**Analysis on reservoir activation with the nonlinearity harnessed from solution-processed molybdenum disulfide**

Songwei Liu,[1] Yingyi Wen,[1] Jingfang Pei,[1] Yang Liu,[1,2] Lekai Song,[1] Pengyu Liu,[1] Xiaoyue Fan,[3] Wenchen Yang,[3] Danmei Pan,[4] Teng Ma,[5] Yue Lin,[4] Gang Wang,[5] Guohua Hu[1,*]

[1]Department of Electronic Engineering, The Chinese University of Hong Kong, Shatin, N. T., Hong Kong SAR, China
[2]Shun Hing Institute of Advanced Engineering, The Chinese University of Hong Kong, Shatin, N. T., Hong Kong SAR, China
[3]Centre for Quantum Physics, Key Laboratory of Advanced Optoelectronic Quantum Architecture and Measurement (MOE), School of Physics, Beijing Institute of Technology, Beijing 100081, China
[4]CAS Key Laboratory of Design and Assembly of Functional Nanostructures, and State Key Laboratory of Structural Chemistry, Fujian Institute of Research on the Structure of Matter, Chinese Academy of Sciences, Fuzhou, Fujian 350002, China
[5]Department of Applied Physics, Hong Kong Polytechnic University, Hung Hom, Kowloon, Hong Kong SAR, China

**ABSTRACT**. Reservoir computing is a recurrent neural network designed for approximating complex dynamics in, for instance, motion tracking, spatial-temporal pattern recognition, and chaotic attractor reconstruction. Its implementation demands intense computation for the nonlinear transformation of the reservoir input, i.e. activating the reservoir. Configuring physical nonlinear networks as the reservoir and employing the physical nonlinearity for the reservoir activation is an emergent solution to address the challenge. In this work, we analyze the feasibility of harnessing the nonlinearity from solution-processed molybdenum disulfide ($MoS_2$) for reservoir activation. We fit the high-order nonlinearity, achieved by Stark modulation of $MoS_2$, as the activation function to facilitate implementation of a reservoir computing model. Due to the high-order nonlinearity, the model can achieve long-term synchronization and robust generalization for complex dynamical system regression. As a potential application exploring this ability, we appoint the model to generate chaotic random numbers for secure data encryption. Given this reservoir activation capability, and the scalability of solution-processed $MoS_2$, our results suggest the potential for realizing physical reservoir computing with solution-processed $MoS_2$.

## I. INTRODUCTION.

Reservoir computing is a recurrent neural network consisting of an input layer, a *reservoir*, and an output layer [1-3]. The reservoir is central to the network – by forging the reservoir into a dynamical system through self-iteration, the network can generate complex dynamical patterns to perform regression in, e.g., motion tracking, spatial-temporal pattern recognition, and chaotic attractor reconstruction [4]. To ensure stability during long-term iteration, the reservoir requires nonlinear activation at each time step, where the activation function used to transform the input must possess a sufficient nonlinearity [5]. Given this, however, the long-term iteration can result in intense computational workload [6]. Leon O. Chua in 1985 purposed that *non-periodic oscillations* can be produced by harnessing the nonlinearity of electronic elements [7]. Inspired by this, configuring nonlinear systems as a physical reservoir and activating the reservoir with the physical nonlinearity is an effective solution to address the intense computation issue [4,8].

Semiconductors with nonlinear properties are proven promising for realizing physical reservoir computing [6,8]. Particularly, two-dimensional (2D) materials are of great interest, given their quantum confinements and the convenience in modulating their electronic structures for nonlinear properties [9]. Indeed, state-of-the-art studies have previewed the feasibility of realizing physical reservoir computing with 2D materials [10,11]. Towards physical realization of reservoir computing with 2D materials, solution-processing holds great promise – it allows for a mass production, and the materials are readily adapted to the CMOS processes and printing techniques for device fabrication [12]. However, a practical realization thus far can be challenging, demanding very-large-scale circuits with high-level integration. Therefore, before physical reservoir computing is realized with solution-processed 2D materials, it is critical to study the feasibly to activate the reservoir with the nonlinearity from solution-processed 2D materials. In fact, similar to the algorithmic approaches, to ensure a successful reservoir activation, it is crucial to modulate solution-processed 2D materials effectively for sufficient nonlinearity. In our previous reports [13,14], we demonstrated that Stark effect is effective to modulate the electronic structures of solution-processed 2D materials for high-order nonlinear properties.

*Contact author: ghhu@ee.cuhk.edu.hk

In this work, we analyze the feasibility of activating the reservoir using the high-order nonlinearity. Here we harness a high-order nonlinearity from solution-processed molybdenum disulfide ($MoS_2$). By fitting the nonlinearity as activation function to implement a reservoir computing model, we prove the model can synchronize and generalize for regression of complex dynamical systems. The model is proven robust to varying hyperparameters, crucial for the rapid on-site deployment of reservoir computing. As a potential application, we show the synchronization ability can be applied in generating chaotic time series with true randomness for, e.g., secure data encryption. Given this reservoir activation capability, and the scalability of solution-processed 2D materials, our results suggest the potential of realizing physical reservoir computing with solution-processed 2D materials.

## II. HIGH-ORDER NONLINEARITY FROM SOLUTION-PROCESSED $MoS_2$

We harness the nonlinearity from $MoS_2$, as $MoS_2$ is an architype of 2D materials with robust properties and, particularly, $MoS_2$ in mono-, few-layer forms allow effective Stark modulation of the electronic structures [13-15]. Figure 1a illustrates the Stark effect in few-layer $MoS_2$ in an electric field – the carriers become delocalized and fill in the intra-layer space, forming intra-layer carrier transport layers. As studied in our previous reports [13,14], applying Stark modulation to solution-processed $MoS_2$, the electrical conductance of the solution-processed nanoflakes and their discrete networks can enhance drastically, leading to a high-order nonlinear electrical property.

Here, we process $MoS_2$ via liquid-phase exfoliation and develop a vertical structure of $MoS_2$ mixed with ferroelectric poly(vinylidene fluoride-co-trifluoroethylene) [P(VDF-TrFE)] polymer for electrical property probing (Fig. 1b inset). See Appendix A and B for the fabrication and electrical characterization. As shown, $MoS_2$ and P(VDF-TrFE) form a composite where the solution-processed $MoS_2$ nanoflakes are embedded in the P(VDF-TrFE) polymer and as such, the ferroelectric P(VDF-TrFE) after polarization can exert local polarization fields to the solution-processed $MoS_2$ nanoflakes and induce Stark modulation [13,14].

Upon P(VDF-TrFE) polarization with external bias, the current output of the structure (i.e. $MoS_2$) shows a nonlinear switching (Fig. 1b). However, hysteresis is associated with the switching due to the polarization hysteresis of P(VDF-TrFE). This can lead to problems with physical reservoir realization – high-frequency data processing by the physical reservoir requires low hysteresis; otherwise, a prolonged time is demanded to forget the previous inputs. To ensure high-frequency data processing, the hysteresis must be minimized. Here we adapt an aging method by cycling with sweeping bias. As charges are injected into and directly pass through the P(VDF-TrFE), the locked dipole states in P(VDF-TrFE) can be easily disrupted, leading to fatigue of the ferroelectric polarization [16]. As observed, fatiguing is demonstrated in the current output – after a few cycles, the hysteretic switching quickly degrades to volatile (Fig. 1c); and after a few more cycles, the hysteresis gradually converges (Fig.

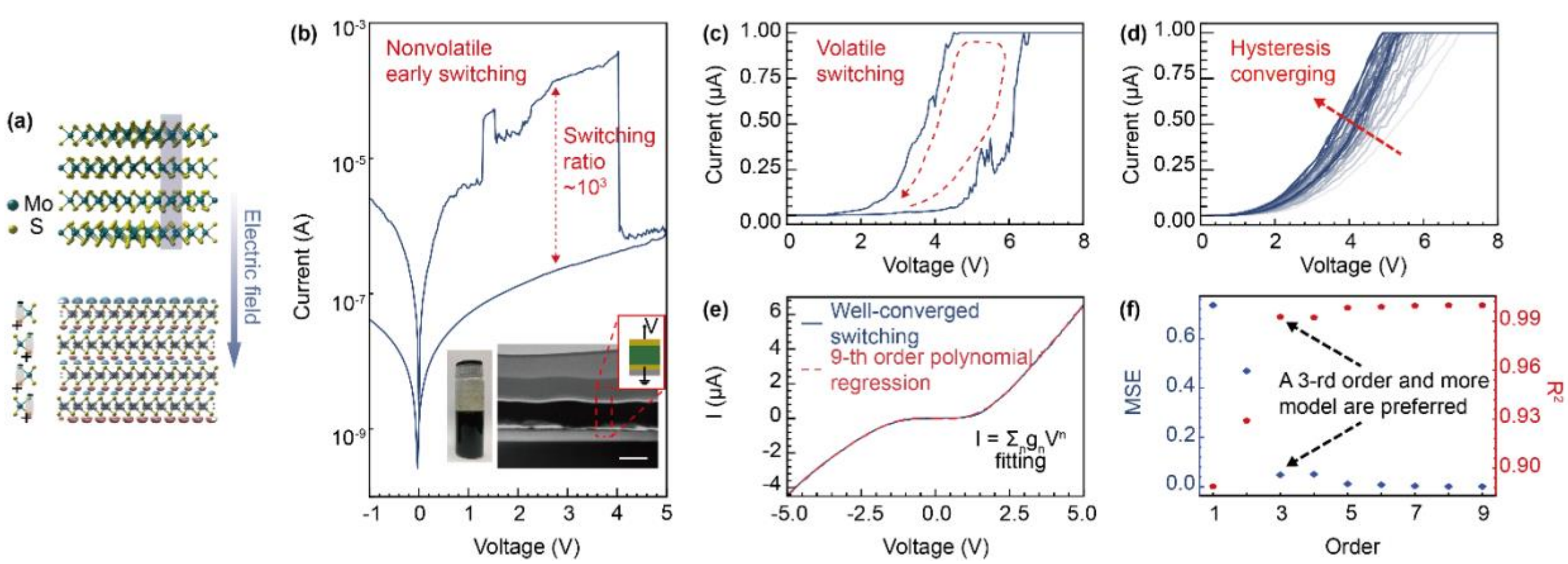

**Figure 1. High-order nonlinearity of solution-processed MoS2.** (a) Schematic formation of conductive two-dimensional carrier transport layers in few-layer $MoS_2$ due to the Stark effect. (b) Initial nonlinear current output with significant hysteresis, with switching ratio of ~$10^3$. Inset – an as-formulated $MoS_2$ ink, and cross-sectional interfacial microscopic image of a typical structure for the tests. The interfacial dark area is the deposited MoS2/ P(VDF-TrFE) layer. Scale bar – 500 nm. (c) Hysteresis fading to volatile switching, and (d) the converging in the switching by fatiguing the P(VDF-TrFE) via bias sweeping. (e) Well-converged nonlinear current output, demonstrating 9-th order polynomial regression fitting. (f) The mean square error (MSE) and coefficient of determination ($R^2$) for the polynomial regression fitting at the different fitting orders.

*Contact author: ghhu@ee.cuhk.edu.hk

1c), leading to well-converged yet nonlinear current output (Fig. 1e).

To quantify the well-fatigued nonlinearity, we use polynomial regression to describe the well-converged current output,

$$I = \sum_{m} g_m V^m + \xi \, , \quad (1)$$

where $m$ is the polynomial order, $\{g_1, g_2, \cdots, g_m\}$ is the corresponding coefficient, and $\xi$ is the residual error. Note that polynomial regression is specifically used here for quantifying the degree of nonlinearity as it can suitably and accurately describe the nonlinearity for the following reservoir activation. As proved in Fig. 1f, only when the polynomial reaches a third order or more, the coefficient of determination ($R^2$) of the fitting is above 0.99 and the mean square error (MSE) is below 0.1. As shown in Fig. 1e, a third order or more quantifies *a high-order nonlinearity* [17].

## III. RESERVOIR ACTIVATION

Solution-processed $MoS_2$ with the proved high-order nonlinearity holds the potential towards the physical realization of reservoir computing. To explore the feasibility, we have implemented a reservoir computing model by fitting the nonlinearity based on RIGEL [18], a physical reservoir computing simulation package developed by our group. See Appendix C for the fabrication and electrical characterization. Figure 2a illustrates the topology of the echo state network (ESN) studied in our work, where the nonlinearity is fitted as the activation function, denoted as $f_{\text{dev}}$, and embedded in the reservoir for nonlinearly transforming the input to the reservoir. Note that echo state network is one of the founding models of reservoir computing [19,20].

The reservoir dynamics of the ESN and its close relative leaky-integrator echo state network (Li-ESN) [21] is described as,

$$\mathbf{r}_n = (1-a)\mathbf{r}_{n-1} + f_{\text{dev}}\left(s_{\text{in}}\widehat{\text{W}}_{\text{in}}\mathbf{p}_n + \rho\widehat{\text{W}}_{\text{res}}\mathbf{r}_{n-1}\right) , \quad (2)$$

where $\widehat{\text{W}}_{\text{in}}$ and $\widehat{\text{W}}_{\text{res}}$ are the weight matrix for the input and reservoir connections, respectively, with $(\mathbf{p}, \mathbf{q})$ representing the input-output pair to the ESN. The coefficient $a$ is the leaking rate, where the primitive ESN is obtained when $a = 1$,

$$\mathbf{r}_n = f_{\text{dev}}\left(s_{\text{in}}\widehat{\text{W}}_{\text{in}}\mathbf{p}_n + \rho\widehat{\text{W}}_{\text{res}}\mathbf{r}_{n-1}\right) . \quad (3)$$

As such, the recurrent relation topology forge the reservoir into a discrete dynamical system, wherein the present reservoir state $\mathbf{r}_n$ is a function of the past states $\{\mathbf{r}_{n-1}, \mathbf{r}_{n-2}, \cdots\}$, with leaking rate $a$, input scaling $s_{\text{in}}$ and reservoir spectral radius $\rho$ as the hyperparameters of the reservoir. This recurrent topology gives the ESN the potential to approximate and forecast time series signals if proper hyperparameter sets are given [3]. $f_{\text{dev}}(\cdot)$ is the activation function, fitted from the high-order nonlinearity of our solution-processed $MoS_2$, to nonlinearly transform the inactivated reservoir input states, $\mathbf{u}_n = s_{\text{in}}\widehat{\text{W}}_{\text{in}}\mathbf{p}_n + \rho\widehat{\text{W}}_{\text{res}}\mathbf{r}_{n-1}$. Figure 2b plots the nonlinear transformation using $f_{\text{dev}}(\cdot)$. Note that $f_{\text{dev}}(\cdot)$ with a limited nonlinearity can result in linear superpositions of the past input states and as such, explosion, and convergence of the reservoir states.

The topology of the ESN family determines its suitability to time series regression. As dynamical systems are in general time series with hidden dynamical patterns, we adopt a weather model proposed by Lorenz in 1963 [22] as the target dynamical system, denoted as *Lorenz-63*, to evaluate the dynamical system regression performance of our ESN. By learning dataset $S = \{P_{\text{train}}, Q_{\text{train}}\} = \{(\tilde{\mathbf{p}}_1^{\text{T}}, \tilde{\mathbf{p}}_2^{\text{T}}, \cdots, \tilde{\mathbf{p}}_N^{\text{T}}), (\tilde{\mathbf{q}}_1^{\text{T}}, \tilde{\mathbf{q}}_2^{\text{T}}, \cdots, \tilde{\mathbf{q}}_N^{\text{T}})\}$ extracted from *Lorenz-63*, the output connection weight $\widehat{\text{W}}_{\text{out}}$ can be optimized by the Tikhonov regularization [23],

$$\widehat{\text{W}}_{\text{out}} = \text{Q}_{\text{train}}\text{R}_{\text{train}}^{\text{T}}\left(\text{R}_{\text{train}}\text{R}_{\text{train}}^{\text{T}} + \alpha^2\hat{\text{I}}\right)^{-1} , \quad (4)$$

where $\text{R}_{\text{train}} = (\mathbf{r}_1^{\text{T}}, \mathbf{r}_2^{\text{T}}, \cdots, \mathbf{r}_N^{\text{T}})$ is the reservoir state matrix generated by the reservoir during the training phase, $\hat{\text{I}}$ is the identity matrix with a dimension of $N \times N$, and $\alpha$ is the regularization factor. During the training phase, the reservoir is left untrained, and only the output connection weight $\widehat{\text{W}}_{\text{out}}$ is optimized. As demonstrated by the training results in Fig. 2c and e, our ESN model successfully learns the hidden dynamics of *Lorenz-63*. For example, during the training phase, corresponding to time step 0 to 3,000, the absolute deviation between the target ground truth and the ESN prediction at each time step $n$, $error(n) = \| \tilde{\mathbf{q}}_n - \mathbf{q}_n \|_2$, remains nearly zero, proving a successful replication of *Lorenz-63*. This demonstrates that our ESN has learned the dynamics of *Lorenz-63* successfully and sufficiently.

In the forecasting phase, using the output connection weight for projecting the high-dimensional reservoir states to the real space where *Lorenz-63* locates, $\boldsymbol{q}_n = \widehat{\text{W}}_{\text{out}}\mathbf{r}_n$, we aim to predict the evolutionary trends of *Lorenz-63*. As demonstrated by the forecasting results in Fig. 2d and e, the ESN successfully predicates the dynamics of *Lorenz-63*. Following the training phase, i.e. beyond time step 3,000, the ESN enters the forecasting phase to predict the evolutionary trends of *Lorenz-63*. As observed, in the short term from time step 3,000 to 3,700, the ESN precisely predicts the *Lorenz-63* evolution. However, as it goes beyond time step 3,700, the ESN starts to drift from the ground truth, and the absolute deviation starts to rise. Nevertheless, the precise short-term prediction of the ESN, in addition to the successful replication of *Lorenz-63*, proves successful activation of the reservoir by using the activation function, $f_{\text{dev}}(\cdot)$, fitted with the high-order nonlinearity from the solution-processed $MoS_2$.

*Contact author: ghhu@ee.cuhk.edu.hk

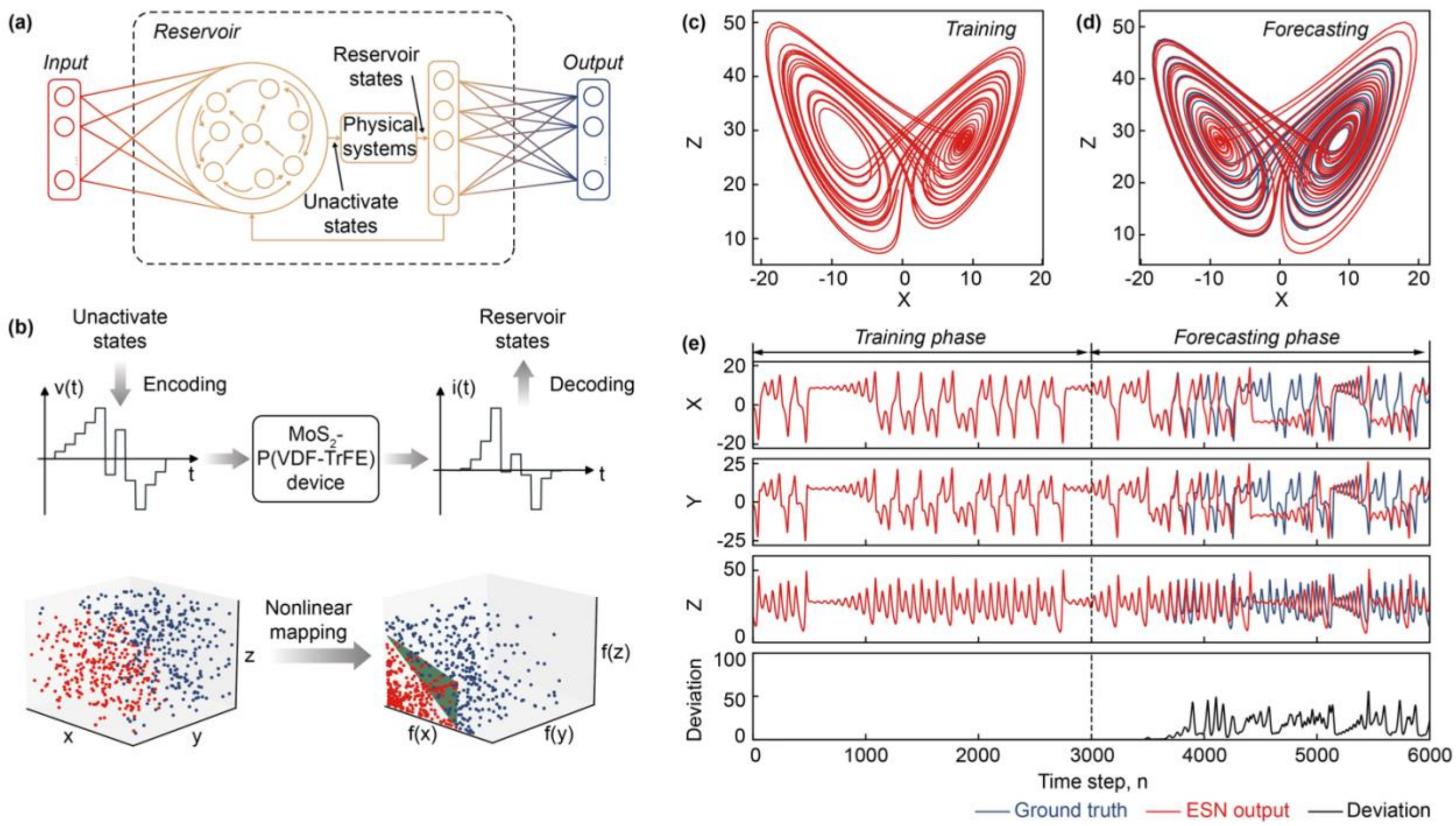

**Figure 2. Reservoir activation.** (a) Topology of the echo state network (ESN), where the reservoir employs the activation function fitted by harnessing high-order nonlinearity from solution-processed $MoS_2$. (b) Schematic reservoir state activation, where the states are nonlinearly mapped by the fitted activation function for activation. Performance of the network in Lorenz-63 approximation task: (c, d) X-Z portraits during the training and forecasting phases; (e) X, Y and Z time series plots, and the corresponding real-time absolute deviation of the network output from the ground truth.

Note that the electrical performance from the solution-processed $MoS_2$ can deviate and exhibit randomness. Based on cyclic measurement at an input voltage of 4 V (Fig. S1), we can see the deviation follows a normal distribution with mean value, $\mu = -1.49$ nA, and standard deviation, $\sigma = 14{,}32$ nA. The reason of this deviation is diverse. For instance, it can originate from the parasitic noise of the characterization instrument or the internal noise of the materials [24]. From this perspective, including the randomness deviation in our ESN model is beneficial. This can be achieved by extending the activation function, $f_{\text{dev}}(\cdot)$, with a random deviation,

$$f_{\text{dev-w-deviation}}(\cdot) = \eta(n) \cdot f_{\text{dev}}(\cdot)\,, \tag{5}$$

where $\eta(n)$ is a random sequence varies along time step $n$, and $\eta(n)$ follows the deviation distribution measured by the electrical characterization. Note that although the ESN model considering the randomness is less effective in forecasting the evolution trend, the long-term synchronization ability of the ESN model is still ensured.

## III. DYNAMICAL SYSTEM SYNCHRONIZATION

Bounded activation functions such as *tanh* and *sigmoid* are often adopted in reservoir computing to ensure asymptotically convergence and continuity of the reservoir states during long-term recurrence, known as the *echo state property* [21,25,26]. Otherwise, the reservoir can explode or converge rapidly [19,21]. To guarantee the *echo state property*, the reservoir dynamics must satisfy the *Lipschitz continuity* [21,27]. In our case, as the fitted activation function, $f_{\text{dev}}$, acts element-wisely on the reservoir inputs, we consider the basic one-dimensional input-output circumstance. Therefore, for two arbitrary reservoir states $r_n$ and $r'_n$,

$$\begin{aligned}
\|r_n - r'_n\|_2 &= \left\| \begin{matrix} (1-a)(r_{n-1} - r'_{n-1}) + \\ f_{\text{dev}}(s_{\text{in}} w_{\text{in}} p_n + \rho w_{\text{res}} r_{n-1}) \\ -f_{\text{dev}}(s_{\text{in}} w_{\text{in}} p_n + \rho w_{\text{res}} r'_{n-1}) \end{matrix} \right\|_2 \\
&= \left\| \begin{matrix} (1-a)(r_{n-1} - r'_{n-1}) + \\ \sum_m g_m \begin{bmatrix} (s_{\text{in}} w_{\text{in}} p_n + \rho w_{\text{res}} r_{n-1})^m \\ -(s_{\text{in}} w_{\text{in}} p_n + \rho w_{\text{res}} r'_{n-1})^m \end{bmatrix} \end{matrix} \right\|_2 \\
&\le \left\| \begin{matrix} \|(1-a)(r_{n-1} - r'_{n-1})\|_2 + \\ \sum_m g_m \begin{bmatrix} (s_{\text{in}} w_{\text{in}} p_n + \rho w_{\text{res}} r_{n-1})^m \\ -(s_{\text{in}} w_{\text{in}} p_n + \rho w_{\text{res}} r'_{n-1})^m \end{bmatrix} \end{matrix} \right\|_2 .
\end{aligned} \tag{6}$$

*Contact author: ghhu@ee.cuhk.edu.hk

However, the polynomial current-output as measured (Fig. 1e) seems to violate the bounded condition as required for the *Lipschitz continuity*, as the unbounded current output can increase to infinity. Fortunately, the operational range can be readily confined by adjusting the input scaling $s_{\text{in}}$, as the input to the reservoir is a function of input scaling where $u_n = U(s_{\text{in}}) = s_{\text{in}} w_{\text{in}} p_n + \rho w_{\text{res}} r_{n-1}$. By setting a proper operational range, $u_n \in [V_{\min}, V_{\max}]$, the current output can be effectively confined. As the fitted activation function, $f_{\text{dev}}$, is monotonic, the reservoir states can satisfy *Lipschitz continuity* by the critical condition,

$$\|f_{dev}(V_{max}) - f_{dev}(V_{min})\|_2 \le \varepsilon \cdot \|r_{n-1} - r'_{n-1}\|_2. \quad (7)$$

Thus, the Eq. (6) then follows

$$\begin{aligned} \|r_n - r'_n\|_2 &\le \|(1-a)(r_{n-1} - r'_{n-1})\|_2 + \left\| \sum_m g_m[(V_{\max})^m - (V_{\min})^m] \right\|_2 \\ &= \|(1-a)(r_{n-1} - r'_{n-1})\|_2 + \|f_{\text{dev}}(V_{\max}) - f_{\text{dev}}(V_{\min})\|_2 \\ &\le (1 - a + \varepsilon) \cdot \|(r_{n-1} - r'_{n-1})\|_2, \end{aligned} \quad (8)$$

ensuring the echo state property during long-term recurrence. The activation function considering the random deviation can be proved to satisfy this property, $f_{\text{dev-w-deviation}}(\cdot)$, based on similar deduction procedures.

As a demonstration, we limit the operational range to -4V to 4V, where the nonlinear current output can be well-bounded. In this context, we examine the long-term synchronization ability of our ESN to *Lorenz-63* using the fitted activation function, $f_{\text{dev}}(\cdot)$. Specifically, in the long-term synchronization test, the time step length of the training phase is set to be 1,000 (instead of 3,000) for short-term initiation of the ESN, while the forecasting phase is 10,000 (instead of 3,000). As demonstrated in Fig. 3a-c, by learning the hidden dynamics of *Lorenz-63* for a short-term, the ESN can perfectly reconstruct the long-term climate, the *Lorenz attractor*, of *Lorenz-63*.

The difference in the chaos behavior between *Lorenz-63* and the ESN reconstruction may give insights to the effectiveness of the fitted activation function, $f_{\text{dev}}$, for activating the reservoir. Given an arbitrary discrete dynamical system $\psi_{n+1} = \mathcal{F}(\psi_n)$ with an initial state $(\psi_0)$, the maximal *Lyapunov* exponent (MLE),

$$\lambda(\psi_0) = \lim_{n\to\infty} \frac{1}{n} \sum_{i=0}^{n-1} ln|\mathcal{F}'(\psi_i)|, \tag{9}$$

can be considered as an indicator characterizing the dynamical behavior of the system [28]. Following the method proposed by M. T. Rosenstein et al. [29], we estimate the MLE of *Lorenz-63* and the ESN reconstruction. As demonstrated in Fig. 3d, both the ESN reconstruction (i.e. the network output) and the *Lorenz-63* attractor (i.e. the target ground truth) exhibit positive MLE, proving chaotic behaviors. In addition, the similarity in MLE (1.168 vs. 1.166) indicates successful approximation of the chaotic characteristics of the *Lorenz-63* attractor. This demonstrates a *generalized synchronization property* [30] of our ESN, meaning that the network can perfectly replicate the chaotic attractor. More importantly, as presented in Fig. 3e, the ESN maintains within the operational range of -4V to 4V as set, proving the effectiveness of the fitted activation function, $f_{\text{dev}}(\cdot)$, for long-term recurrence in operation.

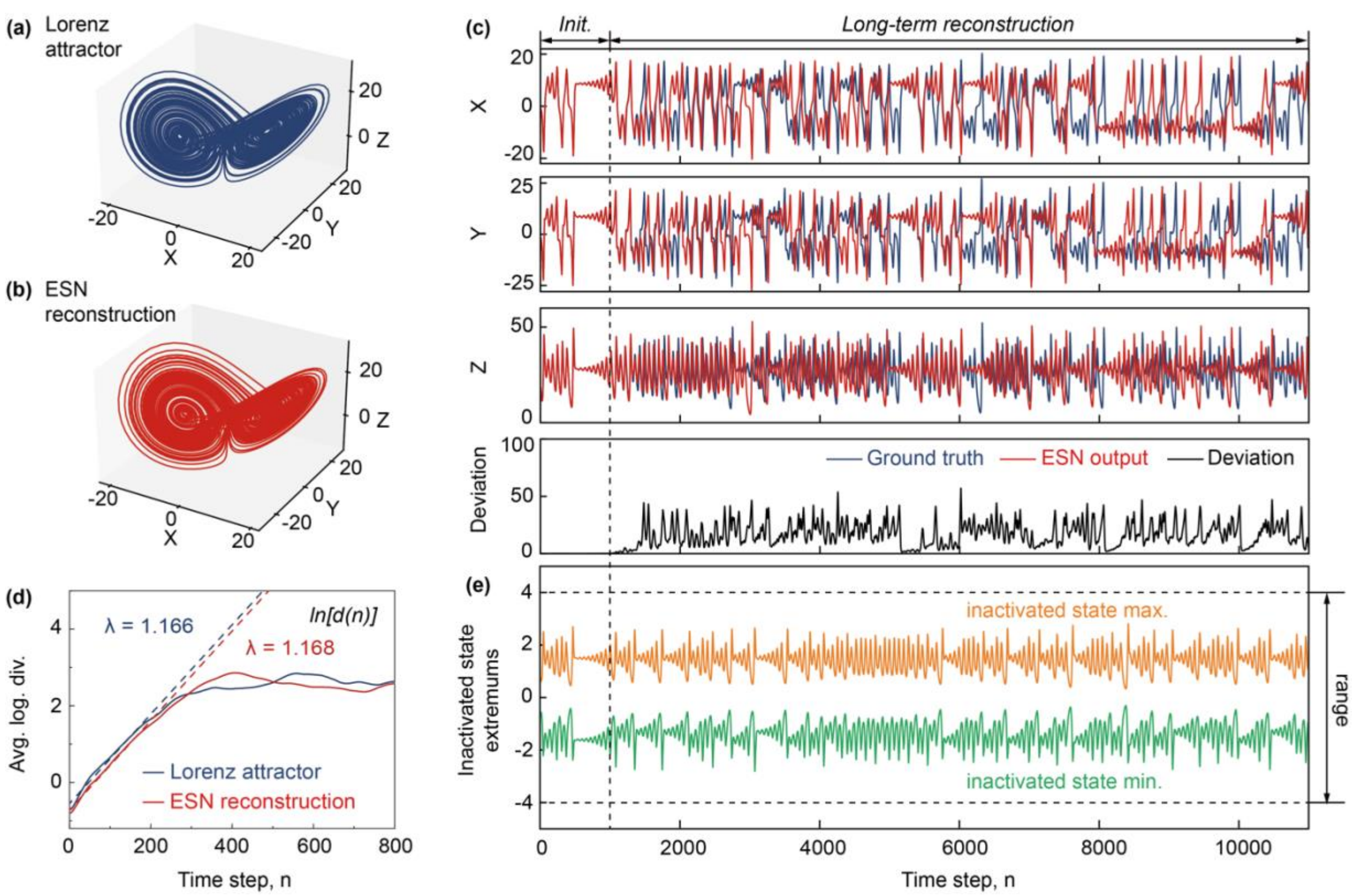

**Figure 3. Long-term synchronization of the ESN.** 3D visualization of (a) the target Lorenz-63 system, and (b) the ESN reconstruction in long-term recurrence. (c) X, Y and Z time series plots in long-term ESN reconstruction, and the corresponding real-time absolute deviation of the network reconstruction from the ground truth. (d) Lyapunov exponent analysis of the target Lorenz-63 system and the ESN approximation, showing a generalized synchronization property of the ESN. The solid and dashed lines are the experimental and fitted curves, respectively. (e) Real-time input inactivated states to the reservoir, showing strict confinement to the operational range from -4 V to 4 V during the long-term recurrence.

*Contact author: ghhu@ee.cuhk.edu.hk

## IV. GENERALIZATION WITH ROBUSTNESS

The generalization ability is crucial for ESN to adapt to the diverse data sets in the different applications. For model-free machine learning, including ESN, capturing the differences in the evolutionary trends of the target systems based on the given dataset is the cornerstone ability of generalization. Back to the dynamical system regression scenario, the different system parameters can lead to varying evolutionary trends. Whether our ESN is capable of capturing the evolutionary trends reflects its generalization ability and also the effectiveness of our approach of using the nonlinearity for reservoir activation. Here, we adopt a single parameter controlled 2D logistic map,

$$\begin{cases} x_{n+1} = h(3y_n + 1) \cdot x_n \cdot (1 - x_n) \\ y_{n+1} = h(3x_{n+1} + 1) \cdot y_n \cdot (1 - y_n) \end{cases}, \tag{10}$$

as the target system to closely examine the generalization ability of our ESN. In general, as shown in Fig. 4a, as the control parameter $h$ increases, the system evolutionary behavior transits from convergence to periodic oscillation and, ultimately, a chaotic behavior. For example, as shown by the system evolutionary phase portrait in Fig. 4b, when $h$ is 1.13, the stabilized target exhibits periodic oscillations, and the ESN easily learns this behavior. When $h$ increases to 1.16, the target is near the bifurcation point, also known as *edge-of-chaos*, and begins to oscillate back and forth between several steady states; the ESN effectively learns the jumping state of the target. Note that due to the synchronization ability of the ESN, instead of perfectly replicating the target, the ESN predicts a higher number of steady states, including the ground truth ones. Eventually, when $h$ reaches 1.19, the target enters chaos, and the ESN perfectly predicts the chaotic oscillation of the target. As shown in Fig. S2 and 3, we then investigate the learning ability of the ESN for dynamical systems of varying dimensions from 1D to 3D. As summarized in Table. SI, the results prove the excellent performance of the ESN in chaotic pattern replication for dynamical systems with varying dimensions. In all cases, the ESN successfully reconstructs the chaotic time series and presents similar MLE. This proves that using the activation function, $f_{\text{dev}}(\cdot)$, fitted from the high-order nonlinearity of our solution-processed $MoS_2$ enables the generalization ability.

As discussed, the reservoir itself is a dynamical system, where varying the hyperparameter sets can impact the evolutionary trends of the reservoir. Although proper output connection weight optimized by Tikhonov regularization can help project rich and diverse reservoir dynamics to the target systems, there are still boundaries. The abundance of effective hyperparameter sets, or alternatively the robustness of the activation functions to the hyperparameter sets, determines the difficulty in setting up a functionable ESN for the specific applications. Specifically, an overly stringent hyperparameter feasible region means that the network requires more refined and complicated optimization methods; on the contrary, a wider range of hyperparameter choices means a stronger stability of the network. To examine the robustness of the fitted activation function, $f_{\text{dev}}$, to the hyperparameters, we appoint the ESN to learn the NARMA-10 task. Note that NARMA-10 is a commonly used benchmark for testing the performance of recurrent neural networks [31]. The NARMA-10 task series can be generated by [32]:

$$\tilde{y}_{n+1} = 0.3\tilde{y}_n + 0.05\tilde{y}_n \left[\sum_{i=0}^{9} \tilde{y}_{n-i}\right] + 1.5v_k v_{k-1} + 0.1, \tag{11}$$

where the input $v$ is a random scalar input drawn from the uniform distribution in the interval [0, 0.5]. As demonstrated in Fig. 4c, the ESN perfectly handles NARMA-10 by replicating the complex dynamics of the ground truth. Note that as shown in Fig. 4d when the input scaling $s_{\text{in}}$ is higher than 3.2, the coefficient of determination $R^2$ between the ground truth and the ESN prediction, $R^2 = 1 - \sum_i (q_i - \tilde{q}_i) / \sum_i (\tilde{q}_i - \langle \tilde{q} \rangle)$, reaches 0.8 and higher. This aligns with our discussion on the input scaling and operational range – a higher input scaling means a wider operational range and leads to larger oscillations in the reservoir states. As such, the larger input scaling makes the ESN run at the

*Contact author: ghhu@ee.cuhk.edu.hk

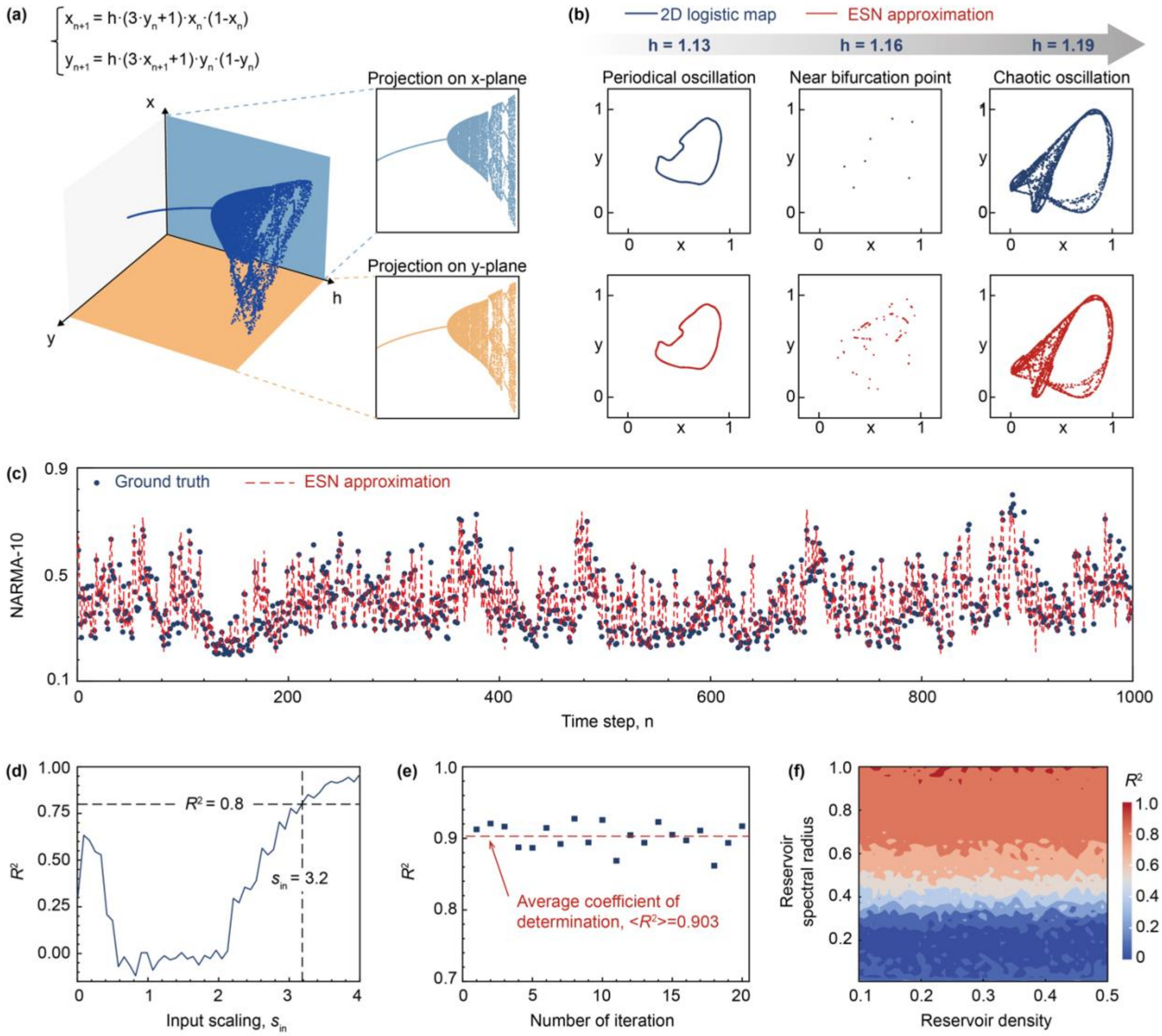

**Figure 4. Generalization ability with robustness to varying hyperparameters.** (a) The bifurcation map of a single parameter controlled 2D dynamical system. (b) The dynamical system forecasting result of the ESN to the system with the different behaviors. The phase portraits shown at $h = 1.13, 1.16$ and 1.19 prove the generalization ability of the ESN. (c) Typical approximation results for the task NARMA-10 by the ESN, showing great agreement between the ESN approximation and the ground truth. (d) The dependency of the network coefficient of determination $R^2$ on the input scaling $s_{in}$, showing the ESN gives credible approximation when the input scaling is over 3.2. (e) The cross-validation of the ESN performance, showing that the ESN achieves an averaged coefficient of determination $R^2$ of over 0.9 at optimized hyperparameters. The results prove robustness of the generalization performance of the ESN. (f) The NARMA-10 performance of the ESN with the different reservoir density and spectral radius. The ESN exhibits good performance ($R^2 \sim 1$) to a wide range of reservoir density and spectral radius combination, showing fitted action function is robust to the hyperparameter variation.

*edge-of-chaos* with a better ability to approximate rich and diverse complex dynamics, meaning the demonstration of a generalization ability.

Our ESN running at the *edge-of-chaos*, however, may not be robust as the fitted activation function, $f_{\text{dev}}(\cdot)$, may not be well bounded and the reservoir may lack negative feedback as a result. To investigate this, we carry out cross-validation examination to forecast NARMA-10 by 20 iterations using $s_{\text{in}} = 4 > 3.2$. As shown in Fig. 4e, the ESN presents robust performance, giving an averaged coefficient of determination $R^2$ over 0.9. Further, we vary the other hyperparameters to test whether robust ESN performance can be maintained. As demonstrated in Fig. 4f, as long as the reservoir spectral radius $\rho$ exceeds 0.5, the ESN guarantees long-term recurrence at the *edge-of-chaos* and generates reliable outcomes, general for the reservoir with a wide range of sparsity (reservoir density ranging from 0.1 to 0.5). The results prove that the activation function, $f_{\text{dev}}(\cdot)$, fitted from the high-order nonlinearity of our solution-processed $MoS_2$ is robust to a wide range of hyperparameter sets, critical for the implementation and the rapid deployment of on-site physical reservoir computing hardware.

*Contact author: ghhu@ee.cuhk.edu.hk

## IV. CHAOTIC NUMBER GENERATION

Our ESN has been proven effective and robust in long-term synchronizing of various dynamical systems. This can be exploited to generate chaotic series by appointing the ESN to learn and imitate a teacher system. The hidden dynamics of the teacher system generated by mathematical algorithms are essentially reproducible, while the chaotic series output by ESN expects to exhibit similar yet differentiated dynamical characteristics to the hidden dynamics, determining that the generated chaotic series are difficult to be reproduced or deciphered.

Chaotic series difficult to be reproduced or deciphered are crucial for research and various engineering problems. For example, chaotic random numbers with true randomness are promising for applications in secure cryptography. For this consideration, we proposed a reservoir computing based chaotic random number generation algorithm (Fig. 5a) and implemented via RIGEL [18]. By selecting a proper teacher chaotic system, the algorithms can successfully generate chaotic times series that is promising for the application in secure cryptography, as shown in Fig. 5

Using the algorithm proposed in Fig. 5, the ESN model can act acts as a chaotic random number generator within an asymmetric encryption strategy illustrated in Fig. 6a, where the hyperparameter sets of the ESN are used as the authentication information to be transmitted by the service provider to the end-users. Particularly, to ensure security, the authentication information alone cannot complete the verification process. Only when both the authentication information and ESN characteristics are available, a proper authentication key can be generated.
To investigate the feasibility of using our ESN in secure cryptography, we test its ability to generate chaotic random numbers. Here, we use a 2D chaotic system as the teacher system,

$$\begin{cases} x_{n+1} = \begin{matrix} \sin\{\pi[4hx_n(1-x_n)] \\ +(1-h)\cdot\sin(\pi y_n)\} \end{matrix} \\ y_{n+1} = \begin{matrix} \sin\{\pi[4hy_n(1-y_n)] \\ +(1-h)\cdot\sin(\pi x_{n+1}^2)\} \end{matrix} \end{cases}, \quad (12)$$

and then appoint the ESN to learn its hidden dynamics for the generation of differentiated chaotic random numbers. As shown in Fig. 6b, the learned ESN generates a set of uniformly yet randomly distributed points within the interval, $\{(x,y)|0 \leq x \leq 1, 0 \leq y \leq 1\}$. See also Fig. 5b for the comparison between the teacher system and the ESN reconstruction. By converting these points into $x-y$ bit pairs, a series of 00, 01, 10, and 11 bit pairs can be generated. Statistical study conducted on 2,500 bit pairs show that the occurrence frequency for each of the bit pairs is ~25% (Fig. 6b), proving a good uniformity. To verify the randomness of the bit stream generated by the ESN, we conduct a standard statistical test of the ESN output using the NIST Special Publication 800-22 test suite [33]. As demonstrated in Fig. 6c, the chaotic random numbers pass 13 out of 15 NIST tests, proving a randomness sufficient for secure cryptography.

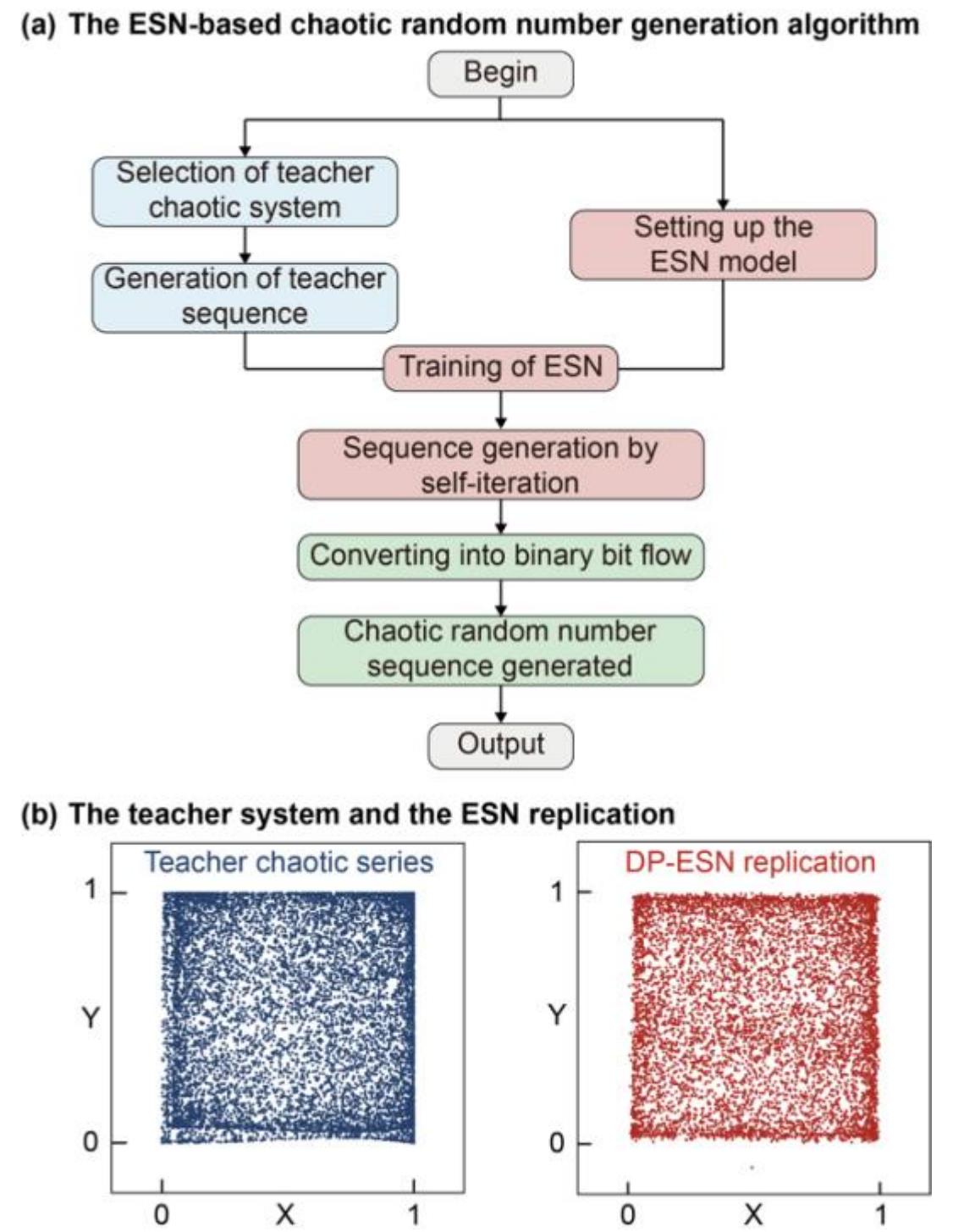

**Figure 5. Chaotic random number generation via the ESN model.** (a) The ESN-based chaotic random number generation algorithm. (b) Comparison between the teacher chaotic time series and the ESN replication.

We perform symmetric-key image encryption using the random bit streams generated by the ESN. The Advanced Encryption Standard (AES) algorithm is used here for encryption and decryption, where CBC and ECB are two representatives of AES algorithms as schematically show in Fig. S4a and b. Using the chaotic random number generated, we successfully perform a symmetric-key image encryption in AES-CBC mode and as observed, the image information is well masked and perfectly restored (Fig. 6d). More explicitly, the uniform color statistical distributions in the three color channels demonstrate that the encrypted image has a uniform distribution throughout the pixel value range from 0 to 255 (Fig. 6e). Meanwhile, the proportion of these three color channels are at the same level, proving that the information is well hidden. The color distributions before and after encryption are also consistent, meaning that the chaotic random numbers generated by the ESN can indeed be used to enable secure cryptography. See also Fig. S4c for the symmetric-key image encryption in AES-ECB mode. The results provide a theoretical foundation of using our ESN in secure cryptography.

*Contact author: ghhu@ee.cuhk.edu.hk

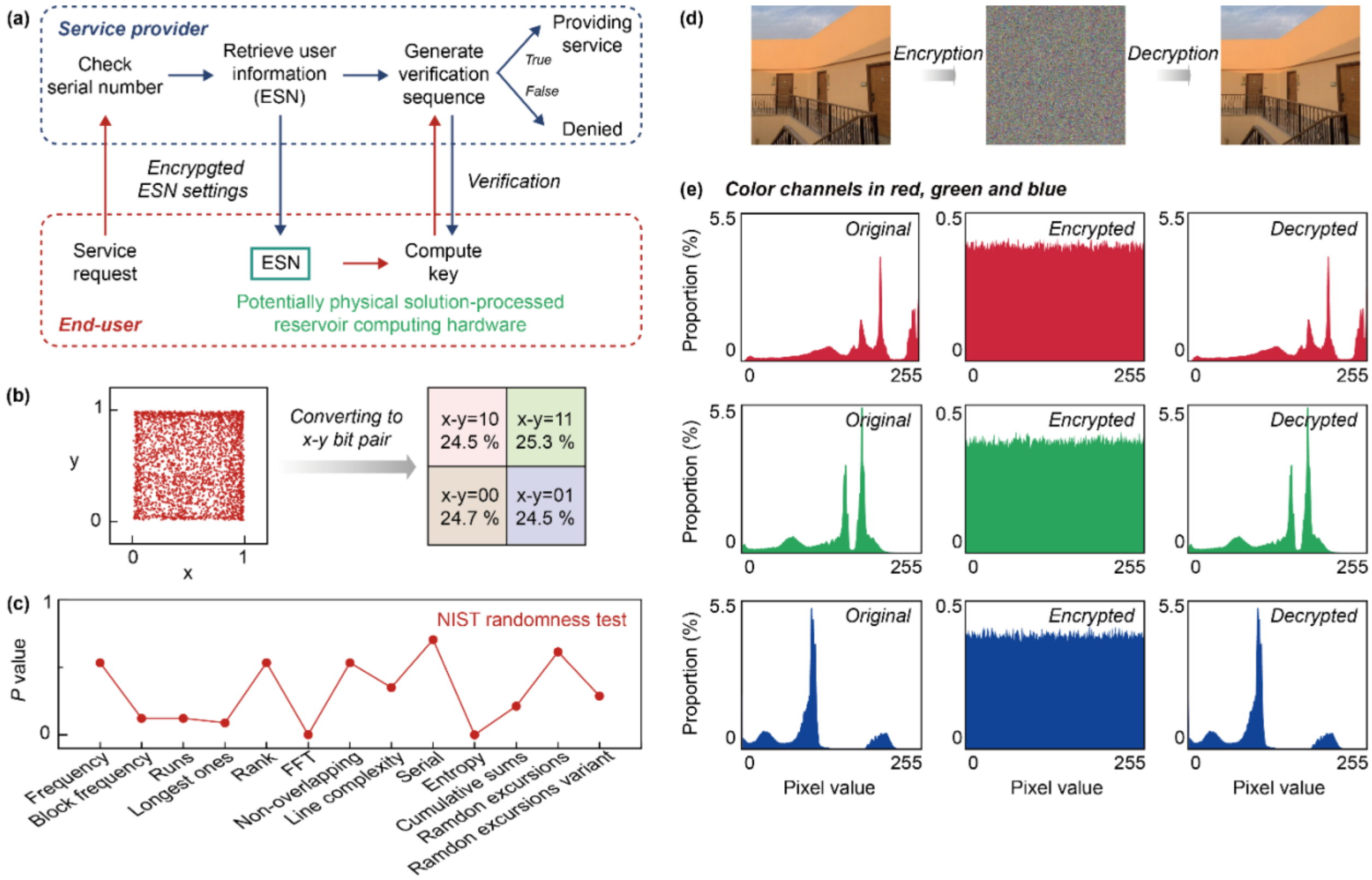

**Figure 6. Application in secure cryptography.** (a) Schematic diagram of authentication scenario between the service provider and the end-user, where the ESN is used as the chaotic random number generator to generate the authentication keys. (b) Schematic illustration of the random bit stream generation via synchronization a chaotic system using the ESN. The left subplot is the phase portrait of the ESN reconstruction result, while the right subplot is the statistical study of the random bit pairs generated by the ESN. The proportion of the 00, 01, 10, and 11 bit pairs are all around 25 %, proving a good uniformity. (c) NIST test result of the chaotic random numbers generated by the ESN, proving the randomness of the chaotic random numbers. (d) Encryption and decryption of an image and (e) the corresponding color distributions using the chaotic random numbers generated by the ESN. The chaotic random number generated by the ESN can effectively mask the information, and the masked information can be perfectly recovered, proving the effectiveness of the ESN random number generator in the application of secure cryptography.

## V. CONCLUSIONS

In this work, we have studied reservoir activation using the high-order nonlinearity harnessed from solution-processed $MoS_2$. The nonlinearity enables activation function fitting for implementing an echo state network model with long-term synchronization and robust generalization. The generalization with robustness is critical for the rapid on-site deployment of physical reservoir computing, while the generalized synchronization to dynamical systems holds promise to enable complex dynamics chasing. The remarkable reservoir activation performance suggests the potential of solution-processed $MoS_2$ for realizing lightweight, efficient physical reservoir computing. The physical reservoir computing implemented can potentially extend the modern edge computing systems for high-efficiency data processing and computing in, for instance, unveiling signal patterns of IoT sensors and healthcare systems, and tracking dynamical motions in wearables and robotics.

## ACKNOWLEDGMENTS

GHH acknowledges support from CUHK (4055115) and NSFC (62304196), YYW and JFP from RGC (24200521), Yang Liu from SHIAE (RNE-p3-21), TM from PolyU (P0042991), Yue Lin from NSFC (52273029), Fujian Science & Technology Innovation Laboratory for Optoelectronic Information of China (2021ZZ119), Pilot Project of Fujian Province

*Contact author: ghhu@ee.cuhk.edu.hk

(2022H0037), Fuzhou Technology Innovation Platform (2022-P-020), Natural Science Foundation of Fujian Province for Distinguished Young Scholars (2023J06045), and Fujian-CAS joint STS Project (2023T3064), and GW from NSFC (12074033), BIT Science and Technology Innovation plan (2022CX01007).

## APPENDIX

### APPENDIX A: SOLUTION PROCESSING $MoS_2$

$MoS_2$ is prepared following our previous reports [13]. Briefly, $MoS_2$ is liquid-phase exfoliated in N-Methyl-2-pyrrolidone (NMP) and then solvent-exchanged into isopropanol/2-butanol (90/10 vol.%) to prepare a 0.5 wt.% inkjet-printable ink. Ferroelectric poly(vinylidene fluoride-co-trifluoroethylene) (70:30) [P(VDF-TrFE)] polymer is dissolved in N,N-dimethylformamide (DMF) to form a 2.5 wt.% inkjet-printable ink. The $MoS_2$ and P(VDF-TrFE) inks are then deposited via inkjet printing by Fujifilm Dimatix Materials Printer DMP-2831 to fabricate the vertical junction structure where $MoS_2$ and P(VDF-TrFE) are sandwiched between the electrodes. The ink cartridge is Dimatix DMC-11610, with a jetting nozzle diameter of 22 μm. The electrodes are deposited by e-beam evaporation (for gold) or magnetron sputtering (for ITO) and patterned by shadow-mask. DMF is purchased from Alfa Aesar, and the other materials are from Sigma-Aldrich. All materials are used as received.

### APPENDIX B: CHARACTERIZATIONS

The current output is measured by Keithley 4200A-SCS. The polarization mapping is measured by atomic force microscope (Brucker Dimension Icon) with a resonant-enhanced piezo-response force microscopy mode.

### APPENDIX C: THE RIGEL

ReservoIr computinG based information sEcurity moduLe (RIGEL) [18] is a physical reservoir computing simulation and data encryption package developed by our group. The two representing member ESN and Li-ESN in the ESN family are implemented. The construction of ESN part refers to the work of Y. Zhao et al. [34].

*Contact author: ghhu@ee.cuhk.edu.hk

*Contact author: ghhu@ee.cuhk.edu.hk

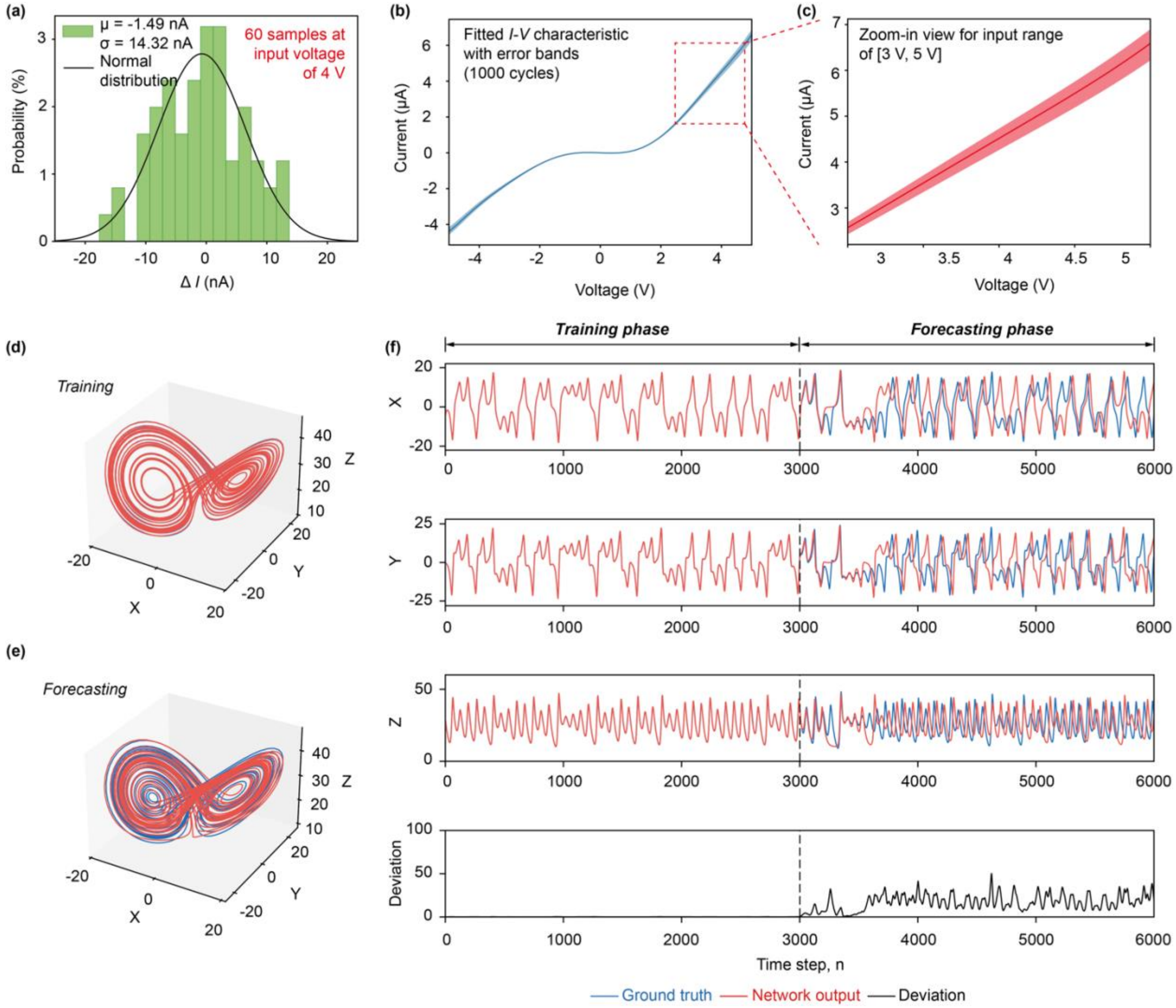

**Figure S1. ESN considering the electrical characterization randomness from the solution-processed $MoS_2$.** (a) The output current deviation statistics by measuring the current deviation from the solution-processed $MoS_2$ and the fitting model at an input voltage of 4 V. The *I-V* characteristics (b) and a zoom-in exhibition (c) of the fitting model considering the randomness gained from (a). (d) – (e) Learning the Lorenz-63 attractor with the ESN considering the randomness. Although the ESN model considering randomness is less effective in forecasting evolution trend, the long-term synchronization ability of the ESN model is still ensured.

*Contact author: ghhu@ee.cuhk.edu.hk

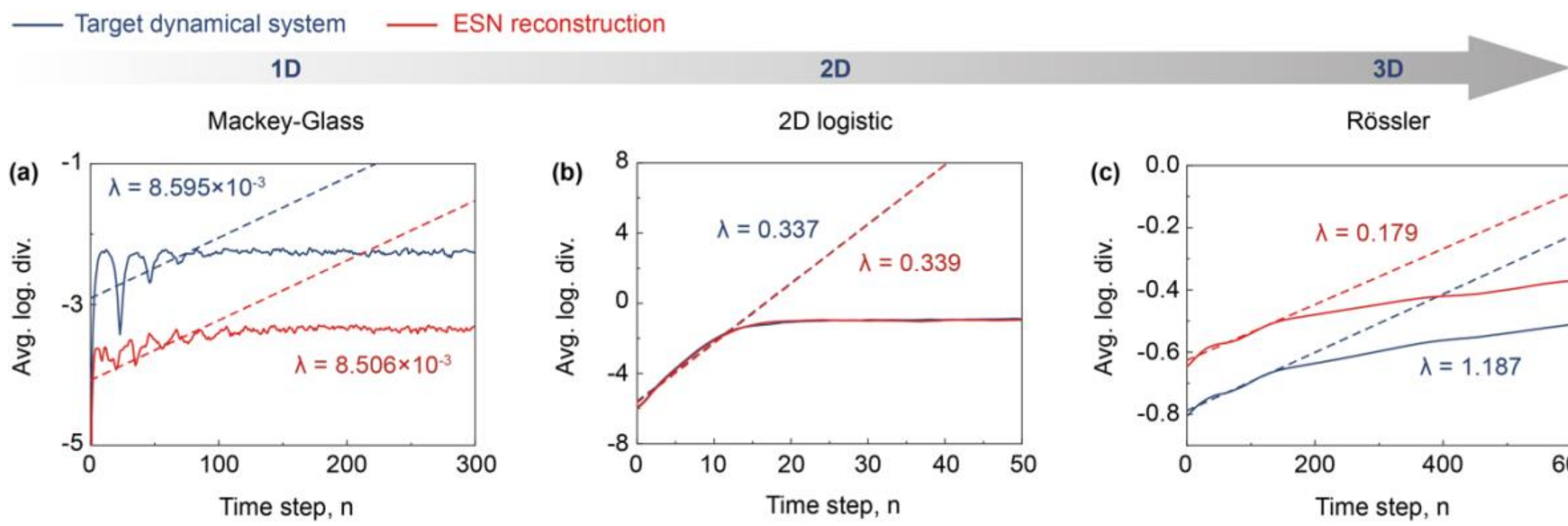

**Figure S2. Maximal *Lyapunov* exponent (MLE) analysis.** The MLE of dynamical systems with different dimensions and the corresponding ESN forecasting results. The target systems – (a) Mackey-Glass (1D dynamical system), (b) 2D logistic (1D dynamical system), and (c) Rössler (3D dynamical system). ESN perfectly estimates the MLE of these three different systems with the different dimensions.

*Contact author: ghhu@ee.cuhk.edu.hk

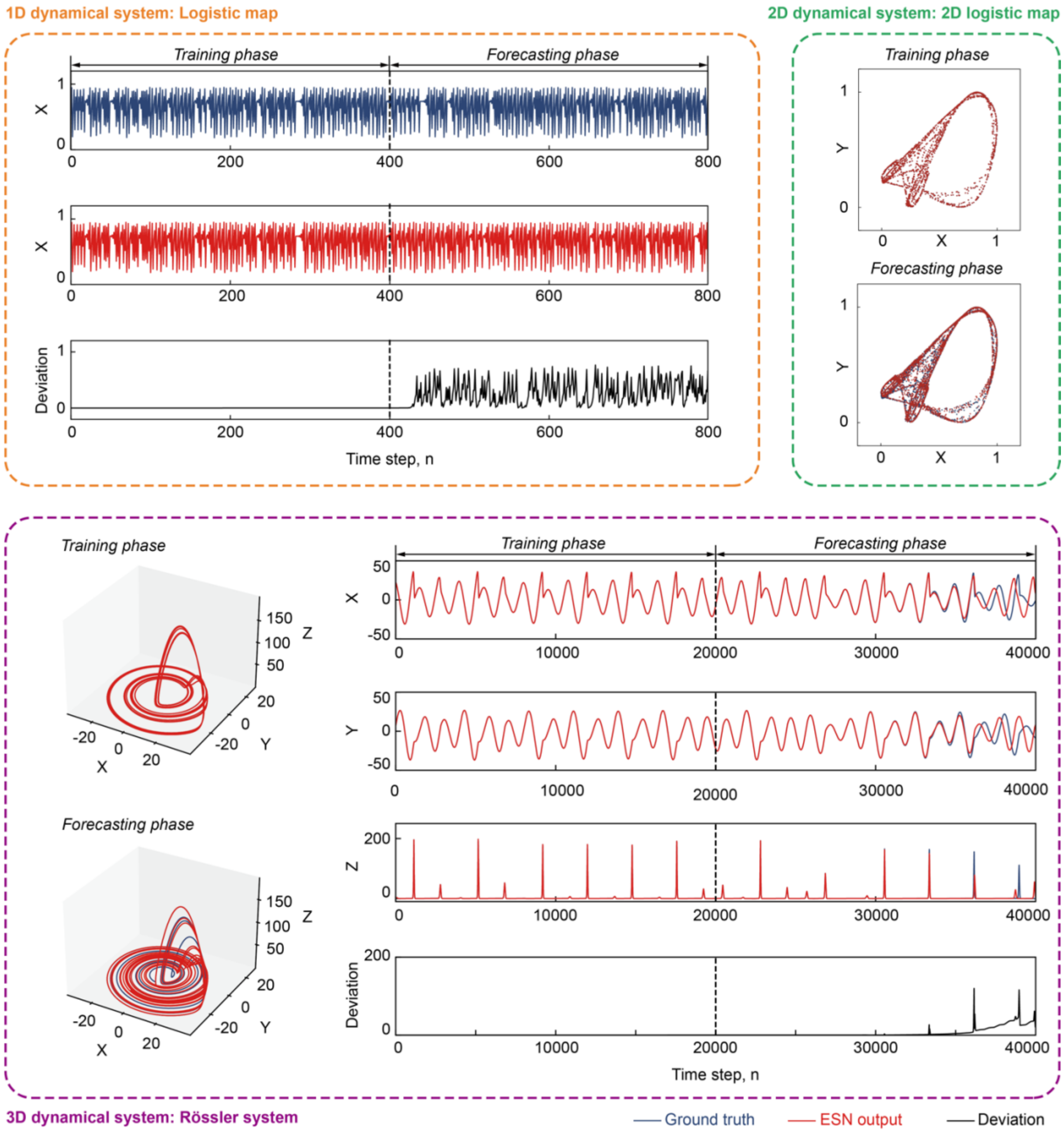

**Figure S3. Performance of ESN with different dimensional dynamical systems.** ESN shows good performance in reconstructing and predicting the chaotic dynamics of these dynamical systems with different dimensions.

*Contact author: ghhu@ee.cuhk.edu.hk

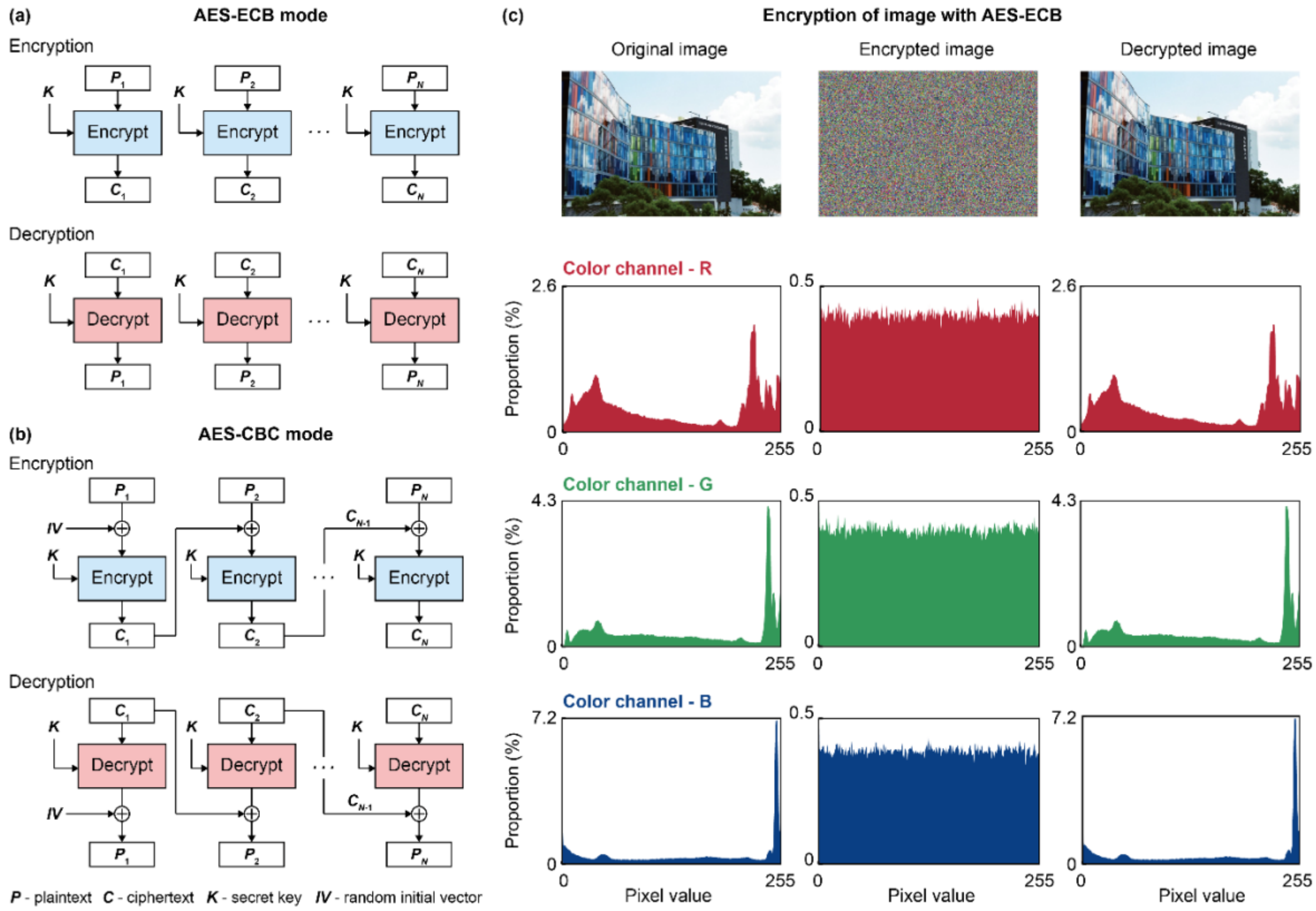

**Figure S4. Data encryption using the random numbers generated by ESN.** Schematic illustration of the AES-ECB (a) and AES-CBC (b) encryption algorithm, respectively. (c) Image encryption using the random numbers generated by ESN. The encryption algorithm used is AES-ECB.

*Contact author: ghhu@ee.cuhk.edu.hk

**Table S1. Chaotic time series approximation and forecasting results by ESN.**

| *System* | *Equations* | *Origin* | *Parameters* | $\Delta t$ | $\lambda$ | $\lambda_{ESN}$ |
|---|---|---|---|---|---|---|
| ***Mackey-Glass*** | $\dot{x} = \frac{b \cdot x(t-\tau)}{1 + x(t-\tau)^a} - c \cdot x(t)$ | $x = 0.5$ | $a = 12$ | 1 | $8.595 \times 10^{-3}$ | $8.506 \times 10^{-3}$ |
| | | | $b = 0.3$ | | | |
| | | | $c = 0.2$ | | | |
| | | | $\tau = 21$ | | | |
| ***2D logistic*** | $x_{n+1} = r(3y_n + 1)x_n(1 - x_n)$ | $x = 0.89$ | $r = 1.19$ | 1 | 0.337 | 0.339 |
| | $y_{n+1} = r(3x_{n+1} + 1)y_n(1 - y_n)$ | $y = 0.33$ | | | | |
| ***Rössler*** | $\dot{x} = -y - z$ | $x = 0.1$ | $a = 0.2$ | 0.005 | 0.187 | 0.179 |
| | $\dot{y} = x + ay$ | $y = 0.1$ | $b = 0.2$ | | | |
| | $\dot{z} = b + z(x - c)$ | $z = 0.1$ | $c = 20$ | | | |
| ***Lorenz-63*** | $\dot{x} = a(y - x)$ | $x = 3$ | $a = 10$ | 0.01 | 1.166 | 1.168 |
| | $\dot{y} = x(b - z) - y$ | $y = 2$ | $b = 28$ | | | |
| | $\dot{z} = xy - cz$ | $z = 16$ | $c = 8/3$ | | | |

*Contact author: ghhu@ee.cuhk.edu.hk